# The Many Facets of Distance and Space:
# the Mobility of Actors in Globally Distributed Project Teams

Tony Clear, Waqar Hussain and Stephen G. MacDonell

*SERL, School of Computing & Mathematical Sciences*
*Auckland University of Technology*
*Private Bag 92006, Auckland 1142, New Zealand*
tony.clear@aut.ac.nz, waqar.hussain@aut.ac.nz, stephen.macdonell@aut.ac.nz

**Abstract**

Global software development practices are shaped by the challenges of time and 'distance', notions perceived to separate sites in a multi-site collaboration. Yet while sites may be fixed, the actors in global projects are mobile, so distance becomes a dynamic spatial dimension rather than a static concept. This empirical study applies grounded theory to unpack the nature of mobility within a three site globally distributed team setting. We develop a model for mapping the movements of team members in local and global spaces, and demonstrate its operation through static snapshots and dynamic patterns evolving over time. Through this study we highlight the complexity of 'mobility' as one facet of 'space' in globally distributed teams and illuminate its tight coupling with the accompanying dimensions of accessibility and context awareness.

## I. INTRODUCTION

This paper reports findings from an in depth empirical study which has employed a grounded analysis of the many dimensions of space in action within a global team setting, across three globally distributed sites. This analysis of a rich body of data expands upon an earlier study [1] which similarly investigated the multi-dimensional nature of 'time' in globally distributed teams (GDTs), by demonstrating the operation and impact of time at the micro level within a distributed educational team setting. As observed in the prior study *"time and space logically belong together in a fuller analysis"* [1]. This paper by contrast investigates the equally rich dimensions of 'space' although again analytically separated from its companion of 'time', and should be viewed as complementary to the previous study.

We first frame this work through a perspective on the concept of 'space' in the context of global teams and global software development (GSD). A necessarily brief coverage of the context of the study and the method of grounded analysis employed, leads to the core of the paper. We outline development of a model of space and mobility, applicable within this global context, progressively abstracted from the grounded data. The model is designed to depict both static and dynamic views. Illustrative examples are presented to demonstrate how emerging constellations of mobility across a number of physical and virtual spatial dimensions may be visually mapped.

These mappings provide new insights into how mobility and space function together in GDTs, and suggest that their inter-operation is richer than understood to date. We argue that this model may be fruitfully applied within GSD contexts to help managers and researchers distinguish between productive and unproductive patterns emerging within such teams. The paper concludes with suggestions for tool support, and raises some questions for future research into the dimensions of mobility and space for actors in global team contexts.

## II. THE QUESTION OF SPACE

Space is a somewhat under-theorized term in globally distributed virtual teams, where spatio-temporal dimensions present themselves daily through the 'challenges of time and distance'. Modernity has been characterized as "the separation of time from space made possible by the standardization of time across the world" [2]. Yet if 'time' can be separated from 'space' in neatly distinct time-zones, then what of space itself and its characteristics, has it become lost in the focus on time? We address the spatial dimension here therefore as a topic of special interest for global virtual teams and in the GSD context.

In their discussion of the phenomenon of 'distance' Olson and Olson identified "four key concepts:

- Common ground.
- Coupling (dependencies) of group work.
- Collaboration readiness - the motivation for coworkers to collaborate.

▪ Collaboration technology readiness - the current level of groupware assimilated by the team'' [3].

At first glance these concepts might seem themselves somewhat distant from any notion of 'space', but perhaps they hint at the innate complexities arising from the physical separation of groups. More recently Carmel and Abbot exploring the notion of 'nearshore' as opposed to 'farshore' note both the significance and the multi-faceted nature of 'proximity'.

> "The customer expects to benefit from one or more of the following constructs of proximity: geographic, temporal, cultural, linguistic, economic, political, and historical linkages" [4].

Nguyen and colleagues working on a more narrow definition of distance conceptualized it as the "Number of sites involved in the communication and completion of a work item" [5]. The findings from their study of the IBM Jazz project ran counter to previous work, "we did not find that geographical distance introduces significant delays in communication and task completion" [5]. They explained this in part by a combination of the culture, practices and technology use engaged in by the team, whose immediate responses to requests and comments from team members, fostered cross site communication, reduced delays, built familiarity and reduced misunderstandings. There may however be unique aspects to this study, the unifying impacts of both IBM corporate culture and the sophistication of a team engaged in building collaborative software such as that in [6]. These practices though echo the strategies recommended by Sarker in addressing problems related to 'space' arising from the three broad categories of: 1) geographical separation, 2) different cultural contexts and 3) different Information Systems Development contexts [5]. The Jazz team practices also appear to have obviated the issues identified by Cramton [7] where lack of "mutual knowledge" engendered distrust between sites.

A broader conceptual discussion of 'space' can be found in the enquiries of Harrison & Dourish into the distinctions between "space and place" [8]. To somewhat oversimplify the arguments, 'spaces' are conceived as bare arenas of potentiality, which need to be adapted to certain human patterns and needs in order to become 'places'. The implication of virtuality and of electronic spaces in these distinctions between space and place remains open to differing interpretation. A nice depiction of the 'place' dimension of a virtual team is given in figure 1 below:

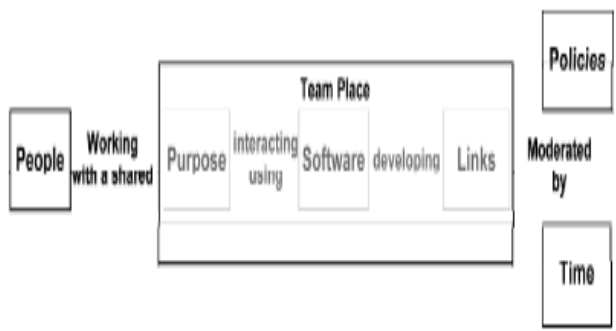

**Figure 1.** The Virtual Team with the place dimension, according to Hamrin and Persson [9, p. 39].

The authors explain thus, "This view on the Virtual Team implies a dual meaning. The first is that People are working with a shared Team Place which is moderated by Policies and Time. The other meaning is that Team Place is created from a shared Purpose, understanding…how to communicate and correctly use Software to form and foster the Links between team members, and thus creating a shared practice. So the Team Place represents the mutual understanding in the team understanding and is shaped by appropriating the Purpose, Software and Links" [9, p.39-40]. Sarker [10] while not reflecting the sophistication of these place and space distinctions, nevertheless frames this in a similar vein by observing that ICTs "replace the physical 'space of place' with the electronic 'space of flows' as the arena for conducting social exchanges".

The final topic to be addressed in this brief survey is the challenge of maintaining 'copresence' [3] both in access to shared objects and in sustaining mutual conversation across sites, an issue not unique to global teams. In their discussion of health professionals and the challenges of developing mobile technologies to support collaboration, Bardram and Hansen [11] note the vital importance of being aware of team colleagues in order to judge how to engage in a cooperative effort. They argue that this "social awareness depends upon knowing the work context of a person" [11]. Technologies that support this such as the 'status' information of instant messenger applications serve to "provide a peripheral and social awareness of fellow workers and friends" [11].

In conclusion, the above discussions of 'space' and 'distance' in GDTs present a complex, but relatively static picture of the actors distributed across geographic and virtual space. This necessarily brief review of the concept of space has introduced some of the innate complexities. Yet it is not until the last paper, (discussing mobile technology support for social awareness in a hospital setting), that the issues associated with the mobility of actors and the implications for technology and collaborative work patterns are addressed. Here we note this omission and highlight the need to investigate the mobility of actors in global teams, to complement the multifaceted views of the phenomenon of 'space' identified from the literature. This very diversity and breadth suggests significant challenges in producing a model for analysing and predicting the facets of 'space' in operation within a GSD context. This study has taken up that challenge.

## III. STUDY CONTEXT AND EMPIRICAL DATA ANALYSIS

The field study reported here (cf. [1, 12] for further detail), investigated the actions of the professional participants in an educational global collaboration across three sites, (AUT University, New Zealand; St Louis University, USA; Uppsala University, Sweden), carried out in late 2004.

While this was an educational collaboration during which students from all three sites worked collaboratively to achieve a common goal, the focus of the study was more specifically on those involved in coordinating and supporting the global collaboration, and in particular their roles and activities of "technology-use mediation" (TUM) [12, 13]. Namely, how they established the technology for the collaboration, how they reinforced and adjusted patterns of use and how they periodically undertook considered major revisions of the supporting technology platforms. Thus while software designer and developer may have been inherent roles for those establishing the technology platform (a custom Lotus Notes application reinforced by the standard AUT University virtual learning environment), the context was more of a GDT of educational and IT professionals than of global software developers.

That said, the issues encountered here are applicable to a wide range of GDTs. They are very much relevant to distributed software teams, where equivalents can be found to many of the 37 independently coded roles identified in this project (e.g. coordinators, team leaders, system support consultants, testers, configurers, trainers and offshore technical coordinators).

The data for the study consisted primarily of a large corpus of email messages spanning more than a year's duration, and covering the phases of the collaboration from inception to completion. This material was complemented by a set of extensive research diary notes, online postings and questionnaire responses, and various documentary artefacts such as: course outlines; instructions to participants; human subject ethics approval documents and assessment guidelines. This presented a large body of textual and digital information for analysis.

The analysis proceeded through a grounded theoretic investigation based upon a 'theoretical sampling' strategy involving selection of specific episodes deemed representative of the four phases of technology-use mediation (establishment, reinforcement, adjustment, and episodic change). An episode of interest was defined as:

> *A relevant temporally bound sequence of events with antecedent conditions and outcomes, which stands apart from others, and has been selected for analysis. [12]*

Eight episodes deemed to be broadly representative were selected for the original study, and codes and concepts were progressively derived by detailed analysis of each episode. This set of eight episodes comprised: one lengthy episode based on a large body of email data (data sources ranged from 1 item to 216 in the largest episode) covering the initial *establishment* phase of the project; four episodes covering *adjustment/reinforcement* TUM activity modes; and three (based mostly on critical incidents) addressing the *episodic change* mode.

Space was but one element of many in the rich collection of codes and concepts that resulted from the analysis, but it was the focus for this extended study into the mobility of actors in GDTs. In this further exploration of the spatial dimensions of the data, the raw data coded under the concept of 'space' was revisited. Here we focus, not on process and detailed codes identified but, on the resulting concepts and relationships. The underlying categories elicited within the concept of 'space' were found to be related to physical/virtual, accessible/inaccessible and indeterminate vs. intermediate spaces. Figure 2 portrays these distinct elements and how they manifested themselves in a concentric structure of three spheres.

The following section elaborates on the spatial categorizations revealed through this grounded analysis, and how they have been adopted in developing a model for mapping mobility in global teams.

## IV. MOBILITY MAPPING MODEL DEVELOPMENT

In this section we present our model which incorporates local and global dimensions, and illustrates potential trajectories of movement for actors traversing boundaries between the various forms of space identified above. In the obverse case we show how actors may become stalled within the three spatial spheres delineated in figures 2 and 3.

### A. Categorization of Space in GDTs

The model in figure 2 shows a categorization of space by means of three spheres of activity. (In this instance a three site model is depicted, but the model could readily accommodate different numbers of sites reflecting the situation being mapped.) The three small circles represent three physical locations in New Zealand, Sweden and United States of America, locations from which local and global virtual teams are formed.

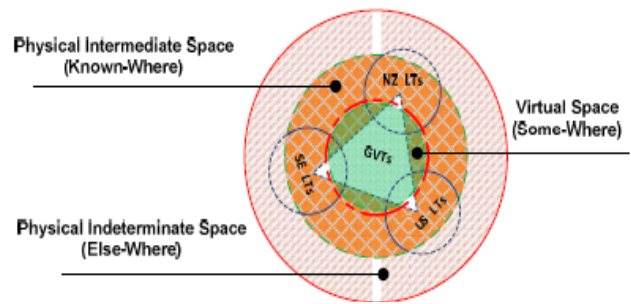

**Figure 2**. Mobility and Spatial Mapping Model for GDTs

### Sphere 3: Physical Indeterminate Space (*Else-Where*)

The specific whereabouts of members in Sphere 3 are undetermined, but these members are generally considered inaccessible for collaboration activity by other members (e.g. on a holiday, on a break, at a conference or just out of the project) unless they <u>choose</u> to connect with the virtual team and announce their availability via some collaborative technology. The outermost circle (Sphere 3) has a solid outer boundary delimiting the extent of global movement by a team member within that sphere. Sphere 2 and Sphere 1 have dotted boundary lines as membranes showing that virtual team members may traverse through these spaces.

### Sphere 2: Physical Intermediate Space (*Known-Where*)

The physical whereabouts of members in this sphere are known to virtual team members, but again they may be accessible or inaccessible, depending on their connection with others via a collaborative technology or platform. From Sphere 2 they may become available for a collaborative activity with members at other sites by crossing the membrane into the virtual space of Sphere 1.

**Sphere 1: Virtual Space (*Some-Where*)**

The whereabouts of members in this space are both physical and virtual, the latter through online presence or traces. This inner most sphere of 'Virtual Space' may be accessible or inaccessible. This is the space where the virtual team members can have (global) access to other members, technology, shared artifacts and other resources. Sphere 1 is the composite collaborative platform in place to support the global activity. This sphere may or may not be accessible depending upon the ability of team members to 'connect' to this zone by means of the available collaborative technology (wikis, repositories, version control software, email, phone, forum, video conferencing and so on).

**B. Mobility and Spatial Mapping Model for GDTs**

Figure 2 above represents a snapshot or constellation operative at any sampled populated with team members can illustrate within the different spatial spheres. But since mobility implies dynamic activity, a sequence of such snapshots allows us to visualize the evolution and (in)stability of mobility patterns over time. The constellation of mobility thus portrayed enable us to see a combination of productive and unproductive patterns of mobility and global collaboration evolving over time. In figure 3 below we demonstrate such evolution by exercising the model dynamically drawing upon samples of data from an illustrative episode.

As in figure 2, the small circles in figure 3 represent physical locations in New Zealand, Sweden and the USA from which local and global virtual teams are formed. The red dotted lines show the capability of virtual team members to access the innermost Sphere 1 – the Virtual Space. When represented as a double hard line segment, this shows the closing of the membrane inhibiting movement between spheres (cf. figure 3 below). This can be cause by a technological breakdown or problem which renders members incapable of accessing the Virtual Space (Sphere 1) and performing collaborative activity.

Figure 3 below depicts the mobility of team members during an episode from the original study (*Adjustment/Reinforcement Episode 3* - [12].This episode records an attempt to set up synchronous sessions across the three sites. The first image (reading from left to right) shows a phone call between two local site coordinators. It is displayed as a 'productive' session as two team members successfully accessed the 'virtual space' or collaborative platform and planned to arrange subsequent synchronous sessions.

The second image shows a 'less productive' event where the New Zealand (NZ) coordinator would not access the platform to communicate with students at other locations, due to the educator's perception of the classroom as a 'local' rather than a 'shared' virtual space. By contrast the Swedish coordinator saw it as global classroom. Image 3 again shows a 'less productive' session as the NZ and US sites were inhibited by the extreme time zone differences imposed by three continents, although the Swedish coordinator sought unsuccessfully to initiate a synchronous collaboration. Image 4 shows the 'most productive' ideal scenario of collaboration if a synchronous session were to happen, in response to an 'announcement' posted to the collaborative platform exhorting students to arrange their own GDT synchronous chat sessions. The last image 5 shows 'less productive' activity as even three way asynchronous collaboration was inhibited by the inability to readily establish a global email list. This required individual action on the part of students at remote sites to set up mail forwarding from their personal hotmail account, which would have taken too long and causes to much confusion. Thus in this scenario only the NZ site had ready access to the 'Virtual Space'.

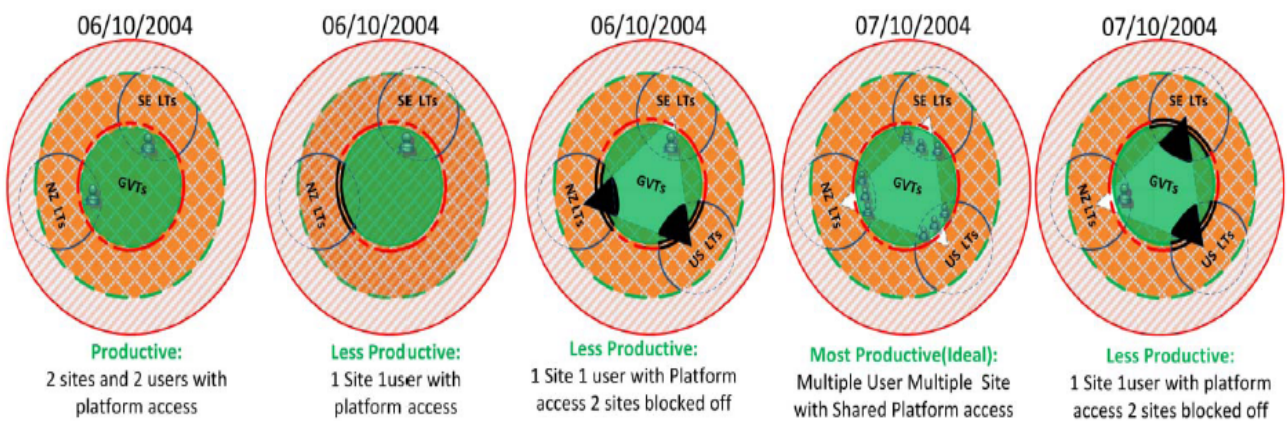

**Figure 3**. Constellation 1 of productive and unproductive patterns of collaboration

In summary the overall episode depicted shows a generally unproductive set of exchanges. The success of the initial two-site phone call was not repeated in the various unsuccessful attempts to link across all three sites (whether synchronously or asynchronously).While not actually realized, image 4 is displayed as the 'most productive' session as it envisage an ideal situation where all team members would successfully access the fully supported and organizationally sanctioned 'Virtual Space' or collaborative platform.

The scenario profiled in figure 3 above demonstrates the ability of this data driven model to show distinct patterns of mobility in a collaborative context. This can be through static snapshots in time or a story told through a dynamically unfolding sequence of events within GDTs.

## V. CONCLUSION

We have presented a model which maps the mobility of actors within three broad spatial spheres present within a globally distributed team contribution of mobility to productive pattern of activity can be portrayed. The insights from this work enable a reinterpretation of the recommendations in [14] for software engineering in a global context:

> The recommendations relating to '*more face to face and collocated meetings and visits*', argue for maximizing the physical space (with '*Physical Intermediate Space*' as the preferred sphere);

> The recommendations for '*collaborative technology infrastructure and synchronous communication*' argue for optimising the '*Virtual Space*', and avoiding the outer loop of '*physical indeterminate space*' and '*inaccessibility*';

> The recommendations for '*lowering task dependencies and using short incremental cycles*', argue for removing the need for common space and mutual awareness by decoupling work. Thus tasks are allocated by strategies such as "sequential" or "parallel segmentation", [15] rather than a task design based upon "tightly coupled work" [3] demanding close collaboration. This recommendation takes advantage of the asynchronous features of the 'Virtual Space' [16] supporting global teamwork through "access to shared objects", while reducing the 'copresence' demands of synchronously "sustaining mutual conversation across sites" [3, 16].

The challenges of space and distance in GDTs may be better overcome through this more sophisticated understanding of space and mobility. The relationships between mobility and mobile computing are obvious linkages to draw, yet the abstract nature of the mobility mapping model presented here may be hard to apply for those holding a strict geographic and locational notion of space. Yet the "definition of realistic mobility models" [17], while critical, is acknowledged as one of the most difficult aspects for designers of systems for mobile environments.

Based upon the richer model of mobility identified here, we highlight the need for further studies to better understand the process of migration between the three spheres of space in operation within distributed teams; typical patterns of mobility for team members within GDTs; and barriers to those movements.

We believe that these insights should prove valuable for managers of global software teams, grappling with the challenges posed by 'time' and 'distance'. We conclude that 'distance' and 'space' may be related (but many faceted) notions, and pervasive within GDTs, but without a deeper understanding of the operation of 'mobility' (which may also represent a form of intersect between space and time), notions of 'space' can only be simplistically viewed.

## ACKNOWLEDGMENT

We wish to thank our colleagues and students at AUT University, Uppsala and St Louis Universities, for their cooperation and support in making this research possible.